\documentclass[aps,prd,reprint,longbibliography,titlepage,nofootinbib]{revtex4-2}

\usepackage{bm}
\usepackage[utf8]{inputenc}
\usepackage[T1]{fontenc}
\usepackage{amsmath}
\usepackage{amsfonts}
\usepackage{mathtools} 
\usepackage{enumerate}
\usepackage{hyperref}
\hypersetup{colorlinks=true,linkcolor=blue,urlcolor=blue,citecolor=blue}
\usepackage[cal=boondox]{mathalfa}
\usepackage{lipsum}
\usepackage{relsize}
\usepackage{color}
\usepackage{nameref}

\newcommand{\intprod}{\mbox{$\;
		\put(0,0){\line(1,0){.9}}\put(.9,0){\line(0,1){1.6}} \; \, \, $}}

\begin{document}
	
	\preprint{AIP/123-QED}
	
	\title{Generalizations of the Nieh-Yan topological invariant}
	
	\author{Merced Montesinos\,\href{https://orcid.org/0000-0002-4936-9170} {\includegraphics[scale=0.05]{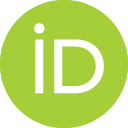}}}
	\email{merced@fis.cinvestav.mx}
	\affiliation{Departamento de F\'{i}sica, Cinvestav, Avenida Instituto Polit\'{e}cnico Nacional 2508, San Pedro Zacatenco,\\
		 07360 Gustavo A. Madero, Ciudad de M\'exico, M\'exico}
	
	\author{Diego Gonzalez\,\href{https://orcid.org/0000-0002-0206-7378} {\includegraphics[scale=0.05]{ORCIDiD_icon128x128.png}}}%
	\email{dgonzalez@fis.cinvestav.mx}
	\affiliation{Departamento de F\'isica de Altas Energ\'ias, Instituto de Ciencias Nucleares, Universidad Nacional Aut\'onoma de M\'exico, Apartado Postal 70-543, Ciudad de M\'exico 04510, M\'exico}
	
	\date{\today}
	
	\begin{abstract}
				We report new topological invariants in four dimensions that are generalizations of the Nieh-Yan topological invariant. The new topological invariants  are obtained through a systematic method along the lines of the one used to get the Nieh-Yan form, but involving an $SO(4,1)$~[$SO(5)$] connection constructed out from an $SO(3,1)$~[$SO(4)$] connection and three $SO(3,1)$~[$SO(4)$] tensor $1$-forms. We give explicit expressions of the new 4-forms that give rise to the new topological invariants for particular choices of these 1-forms and show that the Nieh-Yan form arises as a particular case.
	\end{abstract}

	\maketitle
	
	\section{Introduction}
		
		In the early 1980s, Nieh and Yan derived a four-dimensional topological invariant involving torsion~\cite{Nieh-Yan1982,NiehIJMPA}, which has been since then of great importance in different branches of physics including general relativity~\cite{Date2009,Perez2009,Kaul2012,Sengupta_2013,Corichi2}, $f(\mathfrak{R})$ gravity~\cite{Bombacigno_2021_fR}, and condensed matter physics~\cite{Nissinen_2020_PRL}. Furthermore, there has also been a remarkable interest in exploring the generalizations of the Nieh-Yan topological invariant to higher dimensions. In last direction, we can mention the work of  Mardones and Zanelli~\cite{Mardones_1991} that introduces torsional invariants in higher dimensions and the higher dimensional generalizations of the Nieh-Yan invariant studied by  Chand\'{\i}a and Zanelli~\cite{Zanelli1997}, where explicit expressions for four Nieh-Yan-like invariants in eight dimensions were reported. Subsequently, in Refs.~\cite{Zanelli1998,HanYing1999}  it was shown how to systematically construct Nieh-Yan-like invariants in $4k$ ($k\in Z$) dimensions. Another natural, and perhaps more interesting, direction to generalize the Nieh-Yan topological invariant is to modify its structure but remaining in four-dimensional manifolds. Generalizations of this kind, in particular, would be expected to play a relevant role in the realm of general relativity and related theories of gravity. Recently, it has been reported new torsional topological invariants in four dimensions~\cite{Nieh2018,Montesinos2021}. However, to our knowledge, there are no such four-dimensional generalizations of the Nieh-Yan topological invariant up to now. In this paper we fill out this gap.
		
		More precisely, we present a systematic construction of generalizations of the Nieh-Yan topological invariant in four dimensions. Our approach follows the same spirit of Refs.~\cite{Nieh-Yan1982,Zanelli1997,Nieh2018,Montesinos2021} that leads to the Nieh-Yan invariant and the new torsional invariants in four dimensions recently introduced, but involving an $SO(4,1)$~[$SO(5)$] connection constructed out from an $SO(3,1)$~[$SO(4)$] connection and three $SO(3,1)$~[$SO(4)$] tensor $1$-forms. The new topological invariants, like the Nieh-Yan topological invariant, are integrals of 4-forms that arise from the difference of two Pontrjagin forms. We report explicit expressions for the generalizations of the Nieh-Yan topological invariant and additionally show how the Nieh-Yan invariant emerges as a particular case.

		\section{Preliminaries}
		
		We begin by fixing the notation. In the first-order formalism of general relativity, the fundamental variables to describe the gravitational field are an orthonormal frame of $1$-forms $e^I$ and a connection 1-form $\omega^I{}_J$ that is compatible with the metric $(\eta_{IJ}) =\mbox{diag} (\sigma, 1, 1, 1)$, i.e., $d \eta_{IJ} - \omega^K{}_I \eta_{KJ} - \omega^K{}_J  \eta_{IK}=0$. This means that $\omega_{IJ}=-\omega_{JI}$ since the frame indices $I,J,K, \ldots$ (taking the values $0, 1, 2, 3$) are raised and lowered with $\eta_{IJ}$. The frame rotation group corresponds to $SO(3,1)$ for $\sigma=-1$ and to $SO(4)$ for $\sigma=1$.  The $SO(3,1)$-invariant [$SO(4)$-invariant] tensor $\epsilon_{IJKL}$ is totally antisymmetric and such that $\epsilon_{0123}=1$. The dual basis of $e^I$ is $\partial_I$ and satisfies $\partial_J \intprod \, e^I = \delta^I_J$ where $\intprod$ stands for the contraction of a differential form with a vector field. Furthermore, $R^I{}_J =d \omega^I{}_J + \omega^I{}_K \wedge \omega^K{}_J$ is curvature of $\omega^I{}_J$ and the covariant derivative of an internal tensor that is also a $p$-form $T^{I_1 \cdots I_k}{}_{J_1 \cdots J_l}$ with respect to $\omega^I{}_J$ is given by $D T^{I_1 \cdots I_r}{}_{J_1 \cdots J_l} = d T^{I_1 \cdots I_r}{}_{J_1 \cdots J_l}+ \omega^{I_1}{}_{K_1} \wedge T^{K_1 \cdots I_r}{}_{J_1 \cdots J_l}  + \cdots + \omega^{I_r}{}_{K_r} \wedge T^{I_1 \cdots K_r}{}_{J_1 \cdots J_l} - \omega^{K_1}{}_{J_1} \wedge T^{I_1 \cdots I_r}{}_{K_1 \cdots J_l} - \cdots -  \omega^{K_l}{}_{J_l} \wedge T^{I_1 \cdots I_r}{}_{J_1 \cdots K_l} $. Likewise, the spin connection $\Gamma^I{}_J$ is defined by $d e^I + \Gamma^I{}_J \wedge e^J=0$ and $d \eta_{IJ} - \Gamma^K{}_I \eta_{KJ} - \Gamma^K{}_J  \eta_{IK}=0$, and then $\Gamma_{IJ}=-\Gamma_{JI}$. In addition, $\star$ denotes the Hodge dual operator $\star(e_{I_1} \wedge \dots\wedge e_{I_r}) =(1/(4-r)!) \, \epsilon_{I_1\dots I_rI_{r+1} \dots I_4} e^{I_{r+1}} \wedge \dots \wedge e^{I_4}$ and we define the internal dual of an antisymmetric object $U_{IJ}=-U_{JI}$ as $\ast U_{IJ}:=(1/2) \epsilon_{IJKL} U^{KL}$.

		 Consider the $SO(4,1)$ [$SO(5)$] connection 1-form $\mathcal{A}^a{}_b$ ($\mathcal{A}^{ab}=-\mathcal{A}^{ba}$) constructed out from the connection $\omega^I{}_J$ and the frame $e^I$~\cite{Nieh-Yan1982,Zanelli1997},
		\begin{eqnarray}
		\left ( \mathcal{A}^a{}_b \right ) =
		\left( {\begin{array}{rc}
				\omega^I{}_J  & \frac{1}{l}e^I\\      -\frac{1}{l}e_J & 0     \end{array} } \right ), \label{NYconn}
		\end{eqnarray}
		where $l$ is an arbitrary constant and the indices $a,b,\dots$ run from 0 to 4 and are raised and lowered with the metric $(\eta_{ab}) =\mbox{diag} (\sigma, 1, 1, 1, 1)$. The curvature $\mathcal{F}^a{}_b=d\mathcal{A}^a{}_b + \mathcal{A}^a{}_c \wedge \mathcal{A}^c{}_b$ of $\mathcal{A}^a{}_b$ takes the form
		\begin{eqnarray}
			\mathcal{F}^I{}_J=R^I{}_J-\frac{1}{l^2}e^I\wedge e_J,\hspace{5mm} \mathcal{F}^I{}_4=\frac{1}{l} De^I. \label{NYcurv}
		\end{eqnarray}
		
		It is straightforward to check that the connection $\mathcal{A}^a{}_b$ fulfills the identity
		\begin{equation}
		\mathcal{F}^{a}{}_{b}\wedge \mathcal{F}^{b}{}_{a}=d\left[ \mathcal{A}^{a}{}_{b} \wedge \left( \mathcal{F}^{b}{}_{a} - \frac{1}{3} \mathcal{A}^{b}{}_{c} \wedge \mathcal{A}^{c}{}_{a} \right) \right], \label{PontrF}
		\end{equation}
		which involves the Pontrjagin form $\mathcal{F}^{a}{}_{b}\wedge \mathcal{F}^{b}{}_{a}$. On account of~\eqref{NYconn} and~\eqref{NYcurv}, the identity~\eqref{PontrF} leads to
		\begin{align}
		&R^{I}{}_{J}\wedge R^{J}{}_{I}-\frac{2}{l^2} \left( De^I \wedge De_I - e^I \wedge e^J \wedge R_{IJ} \right) \nonumber \\
		&=d\left[ \omega^{I}{}_{J} \wedge \left( R^{J}{}_{I} - \frac{1}{3} \omega^{J}{}_{K} \wedge \omega^{K}{}_{I} \right) - \frac{2}{l^2} e^I \wedge De_I \right]. \label{PontrF2}
		\end{align}

		Since the parameter $l$ is arbitrary,  we can equate the terms with identical powers of $l$ on both sides of~\eqref{PontrF2}. From the zero power of $l$ we get the identity 
			\begin{equation}
			R^{I}{}_{J}\wedge R^{J}{}_{I}=d\left[ \omega^{I}{}_{J} \wedge \left( R^{J}{}_{I} - \frac{1}{3} \omega^{J}{}_{K} \wedge \omega^{K}{}_{I} \right) \right], \label{PontrR}
		\end{equation}
		satisfied by the Pontrjagin form $R^{I}{}_{J}\wedge R^{J}{}_{I}$, and from the second power of $1/l$ we obtain the identity
		\begin{equation}
			  De^I \wedge De_I - e^I \wedge e^J \wedge R_{IJ} =d\left(e^I \wedge De_I \right), \label{NYiden}
		\end{equation}
		satisfied by the Nieh-Yan form~\cite{Nieh-Yan1982,NiehIJMPA}
		\begin{equation}
			N\equiv De^I \wedge De_I - e^I \wedge e^J \wedge R_{IJ}.\label{NYinv}
		\end{equation}
		 
		 In the next section, we apply the methodology outlined in the previous paragraphs to derive 4-forms that are generalizations of the Nieh-Yan form, whose integrals in four dimensions are topological invariants.

		\section{Generalizations of the Nieh-Yan topological invariant}\label{sec-general}
          
         We begin by defining the $SO(4,1)$ [$SO(5)$] connection 1-form  $\Omega^a{}_b$ ($\Omega^{ab}=-\Omega^{ba}$) by   
         \begin{eqnarray}
         	\left ( \Omega^a{}_b \right ) =
         	\left( {\begin{array}{rc}
         			\omega^I{}_J  & \frac{1}{l}Q^I\\      -\frac{1}{l}Q_J & 0     \end{array} } \right ), \label{conn1}
         \end{eqnarray}
         where $l$ is an arbitrary constant and the $1$-form $Q^I$ is an $SO(3,1)$~[$SO(4)$] tensor whose specific expression is not needed to what follows. The curvature $F^a{}_b=d\Omega^a{}_b + \Omega^a{}_c \wedge \Omega^c{}_b$ of $\Omega^a{}_b$ takes the form
         \begin{eqnarray}
         	F^I{}_J=R^I{}_J-\frac{1}{l^2}Q^I\wedge Q_J,\hspace{5mm} F^I{}_4=\frac{1}{l} DQ^I. \label{curv1}
         \end{eqnarray}
          
         Following the method employed in~\cite{Nieh2018,Montesinos2021}, we further introduce the $SO(4,1)$ [$SO(5)$] connection 1-form  $\Omega'^a{}_b$ defined by
         \begin{equation}
         	\Omega'^a{}_b\equiv\Omega^a{}_b + \xi K^a{}_b,\label{conn2}
         \end{equation}
         where $\xi$ is an arbitrary parameter and $K^a{}_b$  ($K^{ab}=-K^{ba}$) is an $SO(4,1)$ [$SO(5)$] 1-form. Also, as in the case of $Q^I$, the particular expression for $K^a{}_b$ does not need to be specified beforehand. The corresponding curvature of $\Omega'^a{}_b$ is $F'^a{}_b =d \Omega'^a{}_b + \Omega'^a{}_c \wedge \Omega'^c{}_b $ and can be written as
		\begin{equation}
		F'^a{}_b =F^a{}_b + \xi \mathbb{D} K^a{}_b + \xi^2 K^a{}_c \wedge K^c{}_b, 	\label{curv2}
		\end{equation}
    	where $\mathbb{D}$ is the covariant derivative computed with respect to $\Omega^a{}_b$ given by~\eqref{conn1}, and explicitly reads $\mathbb{D} K^a{}_b=d K^a{}_b + \Omega^a{}_c \wedge K^c{}_b - \Omega^c{}_b \wedge K^a{}_c$.
 		
		The identity~\eqref{PontrF} for the connection $\Omega'^a{}_b$ reads
		\begin{equation}
			F'^{a}{}_{b}\wedge F'^{b}{}_{a}=d\left[ \Omega'^{a}{}_{b} \wedge \left( F'^{b}{}_{a} - \frac{1}{3} \Omega'^{b}{}_{c} \wedge \Omega'^{c}{}_{a} \right) \right], \label{PontrFprime}
		\end{equation}
		which involves the Pontrjagin form $F'^{a}{}_{b}\wedge F'^{b}{}_{a}$. Substituting~\eqref{conn2} and~\eqref{curv2} into~\eqref{PontrFprime}, we arrive at 
		\begin{eqnarray}
			&&F^{a}{}_{b}\wedge F^{b}{}_{a} + 2 \xi F^{a}{}_{b} \wedge \mathbb{D} K^b{}_a + \xi^2 \big( \mathbb{D} K^a{}_b  \wedge  \mathbb{D} K^b{}_a  \nonumber \\
			&& \hspace{3mm} + 2 K^a{}_b \wedge K^b{}_c \wedge F^c{}_a \big) + 2 \xi^3 K^a{}_b \wedge K^b{}_c \wedge \mathbb{D} K^c{}_a \nonumber\\
			&& =  d\bigg[ \Omega^{a}{}_{b} \wedge \bigg( F^{b}{}_{a} - \frac{1}{3} \Omega^{b}{}_{c} \wedge \Omega^{c}{}_{a} \bigg)+ \xi \big( \Omega^{a}{}_{b} \wedge \mathbb{D} K^b{}_a \nonumber\\
			&&\hspace{3mm} -\Omega^a{}_b \wedge \Omega^b{}_c \wedge K^c{}_a + K^{a}{}_{b} \wedge F^b{}_a \big) + \xi^2 K^a{}_b \wedge \mathbb{D} K^b{}_a \nonumber\\
			&&\hspace{3mm} + \frac{2}{3} \xi^3 K^a{}_b \wedge K^b{}_c \wedge K^c{}_a \bigg].\label{PontrFprime2}
		\end{eqnarray}
		
			Bearing in mind that $\xi$ is an arbitrary parameter, we can equate the terms with the same powers of $\xi$ on both sides of the identity~\eqref{PontrFprime2}. By doing this, from the zero, first, second, and third powers of $\xi$ we obtain, respectively, the following identities:
			\begin{equation}
			F^{a}{}_{b}\wedge F^{b}{}_{a}=d\left[ \Omega^{a}{}_{b} 	\wedge \left( F^{b}{}_{a} - \frac{1}{3} \Omega^{b}{}_{c} 		\wedge \Omega^{c}{}_{a} \right) \right], \label{Iden-I}
			\end{equation}
			\begin{eqnarray}
				2 F^{a}{}_{b} \wedge \mathbb{D} K^b{}_a &=& d \big(\Omega^{a}{}_{b} \wedge \mathbb{D} K^b{}_a -\Omega^a{}_b \wedge \Omega^b{}_c \wedge K^c{}_a \nonumber \\
				&&+ K^{a}{}_{b} \wedge F^b{}_a \big), \label{Iden-II}
			\end{eqnarray}
		\begin{equation}
			\mathbb{D} K^a{}_b  \wedge  \mathbb{D} K^b{}_a + 2 K^a{}_b \wedge K^b{}_c \wedge F^c{}_a = d \big(K^a{}_b \wedge \mathbb{D} K^b{}_a\big), \label{Iden-III}
		\end{equation}
		\begin{equation}
			K^a{}_b \wedge K^b{}_c \wedge \mathbb{D} K^c{}_a= d \left(\frac{1}{3} K^a{}_b \wedge K^b{}_c \wedge K^c{}_a  \right). \label{Iden-IV}
		\end{equation}
		The identity~\eqref{Iden-I} involves the Pontrjagin form $F^{a}{}_{b}\wedge F^{b}{}_{a}$, whereas the identities~\eqref{Iden-II},~\eqref{Iden-III}, and~\eqref{Iden-IV} involve, respectively, the 4-forms $F^{a}{}_{b} \wedge \mathbb{D} K^b{}_a$, $\mathbb{D} K^a{}_b  \wedge  \mathbb{D} K^b{}_a + 2 K^a{}_b \wedge K^b{}_c \wedge F^c{}_a$, and $K^a{}_b \wedge K^b{}_c \wedge \mathbb{D} K^c{}_a$. The integrals of these 4-forms are topological invariants in four dimensions.
		
		In the following subsections, we use the connection~\eqref{conn1} and its curvature~\eqref{curv1} to write down~\eqref{Iden-I}-\eqref{Iden-IV} and obtain new identities from the different powers of the parameter $l$.
		
		\subsection{Identities coming from~\eqref{Iden-I}}
		
		Substituting~\eqref{conn1} and~\eqref{curv1} into~\eqref{Iden-I}, we obtain
		\begin{align}
			&R^{I}{}_{J}\wedge R^{J}{}_{I}-\frac{2}{l^2} \left( DQ^I \wedge DQ_I - Q^I \wedge Q^J \wedge R_{IJ} \right) \nonumber \\
			&=d\left[ \omega^{I}{}_{J} \wedge \left( R^{J}{}_{I} - \frac{1}{3} \omega^{J}{}_{K} \wedge \omega^{K}{}_{I} \right) - \frac{2}{l^2} Q^I \wedge DQ_I \right].\label{Iden2-I}
		\end{align}
		As before we proceed to match the terms of~\eqref{Iden2-I} with identical powers of $1/l$.  From the zero power of $1/l$ we recover the original identity~\eqref{PontrR} and from the second power of $1/l$ we obtain the new identity
		\begin{equation}
			DQ^I \wedge DQ_I - Q^I \wedge Q^J \wedge R_{IJ} =d\left(Q^I \wedge DQ_I \right). \label{Iden-I-2}
		\end{equation}
		This identity involves the 4-form
		 \begin{equation}
		 	I_1\equiv DQ^I \wedge DQ_I - Q^I \wedge Q^J \wedge R_{IJ}, \label{NewInv1}
		 \end{equation}
		whose integral  in four dimensions is a topological invariant. Notice that $I_1$ constitutes a generalization of the Nieh-Yan form~\eqref{NYinv}. In fact, it is direct to see that in the particular case $Q^I\equiv e^I$, the 4-form~\eqref{NewInv1} reduces to the Nieh-Yan form~\eqref{NYinv}, $I_1=N$. Furthermore, notice from~\eqref{Iden-I} and~\eqref{Iden2-I} that $I_1$ is proportional to the difference of the Pontrjagin forms $F^{a}{}_{b}\wedge F^{b}{}_{a}$ and $R^{I}{}_{J}\wedge R^{J}{}_{I}$. 
				
		As an illustration, let us consider some other particular cases of the 4-form~\eqref{NewInv1}:
		\begin{enumerate}[(i)]
		\item Setting $Q^I\equiv f e^I$ where $f$ is a real function, the 4-form $I_1$ reads
		\begin{eqnarray}
		I_1&=&  D(f e^I) \wedge D(f e_I)  - f^2 e^I \wedge e^J \wedge R_{IJ} \nonumber\\
		&=&f^2\left( De^I \wedge De_I - e^I \wedge e^J \wedge R_{IJ} \right) \nonumber\\
		&&+2f Df \wedge e^I \wedge De_I. \,\,\,\, \label{exampleA1}
		\end{eqnarray}
		Note that this 4-form vanishes if there is no torsion, $De^I=0$. See the conclusions for a comment about the 4-form~\eqref{exampleA1}.
		\item Setting $Q^I\equiv \star(R^I{}_J \wedge e^J)$,  the 4-form $I_1$ takes the form
		\begin{eqnarray}
		I_1&=&D[\star(R^I{}_J \wedge e^J)] \wedge D[\star(R_{IK} \wedge e^K)] \nonumber \\
		&&- \star(R^I{}_K \wedge e^K) \wedge \star(R^J{}_L \wedge e^L)  \wedge R_{IJ}. \label{exampleA2}
		\end{eqnarray}
		Notice that, using $D^2e^I=R^I{}_J\wedge e^J$, the 4-form~\eqref{exampleA2} vanishes when there is no torsion, $De^I=0$.
		\item Setting $Q^I\equiv \star[(\star R^I{}_J) \wedge e^J]=-\sigma \mathfrak{R}^I{}_J e^J$ where $\mathfrak{R}_{IJ}=R^K{}_{IKJ}$ with $R^I{}_J=(1/2)R^I{}_{JKL} e^K \wedge e^L$, the 4-form $I_1$ is
		\begin{eqnarray}
		I_1&=& D(\mathfrak{R}^I{}_J e^J) \wedge D(\mathfrak{R}_{IK} e^K) \nonumber\\
		&&- \mathfrak{R}^I{}_K \mathfrak{R}^J{}_L e^K \wedge  e^L \wedge R_{IJ} \nonumber\\
		&=&\mathfrak{R}^I{}_K \mathfrak{R}^J{}_L \left(\eta_{IJ} De^K \wedge De^L  -e^K \wedge  e^L \wedge R_{IJ}\right) \nonumber\\
		&&-D\mathfrak{R}^I{}_J \wedge D\mathfrak{R}_{IK} \wedge e^J \wedge e^K \nonumber\\
		&& + 2 \mathfrak{R}^I{}_J D\mathfrak{R}_{IK} \wedge De^J \wedge e^K.  
		\end{eqnarray}
		It is worth noting that $\mathfrak{R}_{IJ}\neq \mathfrak{R}_{JI}$ generically because nothing has been stated about torsion.
		\item Setting $Q^I\equiv \star[(\ast R^I{}_J) \wedge e^J]=\sigma G_J{}^I e^J$ where $G_{IJ}=\mathfrak{R}_{IJ}-(1/2) \mathfrak{R} \eta_{IJ}$ with $\mathfrak{R}=\mathfrak{R}^I{}_I$,  the 4-form $I_1$ takes the form 
		\begin{eqnarray}
		I_1&=&  D(G_J{}^I e^J) \wedge D(G_{KI} e^K) \nonumber\\
		&& - G_K{}^I e^K \wedge G_L{}^J e^L \wedge R_{IJ} \nonumber\\
		&=&G_K{}^I G_L{}^J \left(\eta_{IJ} De^K \wedge De^L  -e^K \wedge  e^L \wedge R_{IJ}\right)  \nonumber\\
		&&-DG_J{}^I \wedge DG_{KI} \wedge e^J \wedge e^K \nonumber\\
		&& + 2 G_J{}^I DG_{KI} \wedge De^J \wedge e^K. 
		\end{eqnarray}
		We also remark here that $G_{IJ}\neq G_{JI}$ generically because nothing has been stated about torsion.
		\item Setting $Q^I\equiv \epsilon^{IJKL} \partial_J \intprod \, R_{KL}$, the 4-form $I_1$ reads
		\begin{eqnarray}
		I_1&=&  \epsilon^{IJKL} \epsilon_{IMNP} D(\partial_J \intprod \, R_{KL}) \wedge D(\partial^M \intprod \, R^{NP})\nonumber\\
		&& - \epsilon^{IKLM} \epsilon^{JPQS}(\partial_K \intprod \, R_{LM}) \wedge (\partial_P \intprod \, R_{QS}) \wedge R_{IJ} \nonumber\\
		&=&  2 \sigma    D(\partial_I \intprod \, R_{JK}) \wedge \big[  D(\partial^I \intprod \, R^{JK}) - 2 D(\partial^J \intprod \, R^{IK}) \big]  \nonumber\\
		&& + 2 \sigma  (\partial_I \intprod \, R_{JK}) \wedge \big[ 2 (\partial^J \intprod \, R^{KL}) \wedge R^{I}{}_{L}   \nonumber\\
		&& + (\partial^L \intprod \, R^{JK}) \wedge R^{I}{}_{L}-2 (\partial^I \intprod \, R^{KL}) \wedge R^{J}{}_{L}  \nonumber\\
		&&+ 2 (\partial^K \intprod \, R^{IL}) \wedge R^{J}{}_{L}- 2 (\partial^L \intprod \, R^{IK}) \wedge R^{J}{}_{L}\big].
		\end{eqnarray}
		\end{enumerate}
		
		It is not difficult to realize that two other possible choices for $Q^I$ are $Q^I\equiv \epsilon^{IJKL} \partial_J \intprod \, (\ast R_{KL})$  and $Q^I\equiv \epsilon^{IJKL} \partial_J \intprod \, (\star R_{KL})$. Nevertheless, since $ \epsilon^{IJKL} \partial_J \intprod \, (\ast R_{KL})=-2\sigma \mathfrak{R}^I{}_J e^J$  and $\epsilon^{IJKL} \partial_J \intprod \, (\star R_{KL})=2\sigma G_J{}^I e^J$, they correspond to the cases (iii) and (iv)  and thus there is no need to consider them. It is worth mentioning that the choices for $Q^I$ presented are not meant to be exhaustive, but rather illustrative of the variety of 4-forms that can arise from~\eqref{NewInv1}.

		\subsection{Identities coming from~\eqref{Iden-II}}
		Using~\eqref{conn1} and~\eqref{curv1}, the identity~\eqref{Iden-II} leads to
		\begin{align}
			&2  R^{I}{}_{J}\wedge DK^{J}{}_{I} -\frac{2}{l} \big( 2DK^{I}{}_{4} \wedge DQ_I - K^I{}_4 \wedge Q^J \wedge R_{IJ}   \nonumber\\
			&\hspace{3mm}- Q^I \wedge  K^J{}_4 \wedge R_{IJ} \big) \nonumber\\
			&\hspace{3mm} - \frac{2}{l^2} \big( Q^I \wedge Q_J \wedge DK^J{}_I + 2 DQ^I \wedge Q_J \wedge K^J{}_I  \big) \nonumber\\
			&= d\bigg[ \omega^{I}{}_{J} \wedge DK^{J}{}_{I} - \omega^{I}{}_{J}\wedge \omega^{J}{}_{K}\wedge K^{K}{}_{I} + K^{I}{}_{J}\wedge R^{J}{}_{I} \nonumber\\
			&\hspace{3mm} -\! \frac{2}{l}  \big( DK^{I}{}_{4} \wedge Q_I + K^{I}{}_{4} \wedge DQ_I  \big) \!-\! \frac{2}{l^2} Q^I \wedge Q_J \wedge K^J{}_{I} \bigg]\!. \label{Iden2-II}
		\end{align}
		Next, equating the terms of~\eqref{Iden2-II} with the same powers of $1/l$, from the zero, first, and second powers of $1/l$ we obtain, respectively, the following identities:
		\begin{eqnarray}
			2  R^{I}{}_{J}\wedge DK^{J}{}_{I} &=& d  \big( \omega^{I}{}_{J} \wedge DK^{J}{}_{I}  - \omega^{I}{}_{J}\wedge \omega^{J}{}_{K}\wedge K^{K}{}_{I}  \nonumber\\
			&&+ K^{I}{}_{J}\wedge R^{J}{}_{I} \big), \label{Iden-II-1}
		\end{eqnarray}
		\begin{eqnarray}
			&&2DK^{I}{}_{4} \wedge DQ_I - K^I{}_4 \wedge Q^J \wedge R_{IJ} - Q^I \wedge K^J{}_4 \wedge R_{IJ} \nonumber\\
			&&=d \left( DK^{I}{}_{4} \wedge Q_I + K^{I}{}_{4} \wedge DQ_I  \right) ,\label{Iden-II-2}
		\end{eqnarray}
		\begin{eqnarray}
			&&Q^I \wedge Q_J \wedge DK^J{}_I + 2 DQ^I \wedge Q_J \wedge K^J{}_I \nonumber\\
			&&= d \left( Q^I \wedge Q_J \wedge K^J{}_{I} \right). \label{Iden-II-3}
		\end{eqnarray}
		
		The identity~\eqref{Iden-II-1} involves the 4-form 
		 \begin{equation}
		 	R^{I}{}_{J}\wedge DK^{J}{}_{I},
		 \end{equation}
		 which was first reported in~\cite{Montesinos2021}, whereas the identities~\eqref{Iden-II-2} and~\eqref{Iden-II-3} show, respectively, that the integrals of the 4-forms
		\begin{eqnarray}
			I_2 &\equiv& 2DK^{I}{}_{4} \wedge DQ_I - K^I{}_4 \wedge Q^J \wedge R_{IJ}  \nonumber \\
			&& - Q^I \wedge K^J{}_4 \wedge R_{IJ}, \label{NewInv2}
		\end{eqnarray}
		\begin{equation}
			I_3\equiv Q^I \wedge Q_J \wedge DK^J{}_I + 2 DQ^I \wedge Q_J \wedge K^J{}_I, \label{NewInv3}
		\end{equation}
		are topological invariants. Notice that the 4-form $I_2$ is a generalization of the 4-form~\eqref{NewInv1}, and then, in turn, is a generalization of the Nieh-Yan topological form~\eqref{NYinv}. In fact, if we set $K^{I}{}_{4}\equiv Q^I$, it is straightforward to see that $I_2=2 I_1$.  Clearly, there are several plausible choices for $K^{I}{}_{4}$ and $Q^I$. However here, for illustrative purposes, we limit ourselves to two particular cases of~\eqref{NewInv2}:
			\begin{enumerate}[(i)]
			\item Setting $K^{I}{}_{4}\equiv e^I$ and $Q^I\equiv \star(R^I{}_J \wedge e^J)$,  the 4-form $I_2$ takes the form
				\begin{eqnarray}
				I_2 &=& 2De^I \wedge D[\star(R_{IJ} \wedge e^J)] - e^I \wedge \star(R^J{}_K \wedge e^K) \wedge R_{IJ}  \nonumber \\
				&&- \star(R^I{}_K \wedge e^K) \wedge e^J \wedge R_{IJ}. \label{ExI2}
		    	\end{eqnarray}
	    	Notice that this 4-form vanishes if there is no torsion, $De^I=0$.
		   \item Setting $K^{I}{}_{4}\equiv e^I$ and $Q^I\equiv \star[(\star R^I{}_J) \wedge e^J]=-\sigma \mathfrak{R}^I{}_J e^J$, the 4-form $I_2$ can be written as
		    \begin{align}
			I_2 =& -\sigma \big( 2 \mathfrak{R}_{IJ} De^I \wedge De^J + 2 D\mathfrak{R}_{IJ} \wedge De^I \wedge e^J  \nonumber \\
			& - \mathfrak{R}^J{}_K e^I \wedge e^K \wedge R_{IJ} - \mathfrak{R}^I{}_K e^K \wedge e^J \wedge R_{IJ}  \big).
		   \end{align}
		\end{enumerate}
		To illustrate~\eqref{NewInv3} we also have a wide variety of choices for $Q^I$ and $K^I{}_J$. In particular, it is interesting to note that taking $Q^I\equiv e^I$ and $K^I{}_J\equiv\omega^I{}_J - \Gamma^I{}_J$ (contorsion), the 4-form~\eqref{NewInv3} reduces to the Nieh-Yan form~\eqref{NYinv}, $I_3=N$. This can be checked directly by using $De^I=K^I{}_J \wedge e^J$ and $D^2e^I=R^I{}_J\wedge e^J$. Among the other possible alternatives for $Q^I$ and $K^I{}_J$, we consider here the following two cases:
		\begin{enumerate}[(i)]
			\item Setting $Q^I\equiv \star[(\star R^I{}_J) \wedge e^J]$ and $K^I{}_J\equiv\omega^I{}_J - \Gamma^I{}_J$, the 4-form $I_3$ reads 
			\begin{eqnarray}
				I_3 &=& - \mathfrak{R}^I{}_K \mathfrak{R}^J{}_L e^K \wedge  e^L \wedge DK_{IJ} \nonumber\\
				&& - D(\mathfrak{R}^I{}_K e^K) \wedge \mathfrak{R}^J{}_L e^L \wedge K_{IJ} \nonumber\\
				&=&-\mathfrak{R}^I{}_K \mathfrak{R}^J{}_L \big(e^K \wedge  e^L \wedge DK_{IJ}+2De^K \wedge e^L \wedge K_{IJ}\big) \nonumber\\
				&&-2 \mathfrak{R}^J{}_L D\mathfrak{R}^I{}_K e^K \wedge  e^L \wedge K_{IJ}. 
			\end{eqnarray}
			Since $De^I=K^I{}_J \wedge e^J$, this 4-form vanishes in the absence of torsion, $De^I=0$.
			\item Setting $Q^I\equiv \star(R^I{}_J \wedge e^J)$ and $K^I{}_J\equiv T_{K}{}^I{}_J e^K$ where $De^I\equiv (1/2) T^{I}{}_{JK} e^J\wedge e^K$ with $T^I{}_{JK}=-T^I{}_{KJ}$, the 4-form $I_3$ takes the form
			\begin{align}
				I_3 =& \star(R^I{}_K \wedge e^K) \wedge \star(R_{JL} \wedge e^L) \wedge D\big(T_{K}{}^J{}_I e^K\big) \nonumber\\
				&+ 2 D\big[\star(R^I{}_K \wedge e^K)\big] \wedge \star(R_{JL} \wedge e^L) \wedge \big(T_{K}{}^J{}_I e^K\big).
			\end{align}
		  Clearly, this 4-form vanishes when torsion vanishes, $De^I=0$.
		\end{enumerate}
	
		\subsection{Identities coming from~\eqref{Iden-III}}
		
		Taking into account~\eqref{conn1} and~\eqref{curv1}, the identity~\eqref{Iden-III} leads to
	\begin{align}
		&DK^{I}{}_{J}\wedge DK^{J}{}_{I} + 2 K^{I}{}_{J}\wedge K^{J}{}_{K}\wedge R^{K}{}_{I} + 2 DK^I{}_4 \wedge DK^4{}_I \nonumber\\
		&\hspace{3mm}+ 2 K^I{}_4 \wedge K^4{}_J \wedge R^{J}{}_{I} - \frac{4}{l} \big(K^I{}_4 \wedge Q_J  \wedge DK^{J}{}_{I}   \nonumber\\  
		&\hspace{3mm}+ DK^I{}_4 \wedge Q_J  \wedge K^{J}{}_{I}- K^I{}_4 \wedge DQ_J  \wedge K^{J}{}_{I}  \big)   \nonumber\\
		&= d\bigg( K^{I}{}_{J}\wedge DK^{J}{}_{I} + 2 K^I{}_4 \wedge DK^4{}_I \nonumber\\
		&\hspace{3mm}-\frac{4}{l} K^I{}_4 \wedge Q_J \wedge K^{J}{}_{I} \bigg). \label{Iden2-III}
	\end{align}
	Matching the terms of~\eqref{Iden2-III} with identical powers of $1/l$, from the zero and first powers of $1/l$ we get, respectively, the following identities:
	\begin{align}
		&DK^{I}{}_{J}\wedge DK^{J}{}_{I} \! + \! 2 K^{I}{}_{J}\wedge K^{J}{}_{K}\wedge R^{K}{}_{I} \! + \! 2 DK^I{}_4 \wedge DK^4{}_I \nonumber\\
		&\hspace{3mm}+ 2 K^I{}_4 \wedge K^4{}_J \wedge R^{J}{}_{I} \nonumber\\
		&= d\big( 	K^{I}{}_{J}\wedge DK^{J}{}_{I} + 2 K^I{}_4 \wedge DK^4{}_I \big), \label{Iden-III-1}
		\end{align}
		\begin{eqnarray}
			&&K^I{}_4 \wedge Q_J  \wedge DK^{J}{}_{I} + DK^I{}_4 \wedge Q_J  \wedge K^{J}{}_{I}   \nonumber\\
			&&- K^I{}_4 \wedge DQ_J  \wedge K^{J}{}_{I} = d\left( K^I{}_4 \wedge Q_J \wedge K^{J}{}_{I} \right). \label{Iden-III-2}
		\end{eqnarray}
		The identities~\eqref{Iden-III-1} and~\eqref{Iden-III-2} show, respectively, that the integrals of the 4-forms
			\begin{eqnarray}
			I_4&\equiv&  DK^{I}{}_{J}\wedge DK^{J}{}_{I} + 2 K^{I}{}_{J}\wedge K^{J}{}_{K}\wedge R^{K}{}_{I} \nonumber\\
			&&+ 2 DK^I{}_4 \wedge DK^4{}_I + 2 K^I{}_4 \wedge K^4{}_J \wedge R^{J}{}_{I}, \label{NewInv4}\\
			I_5&\equiv& K^I{}_4 \wedge Q_J  \wedge DK^{J}{}_{I} + DK^I{}_4 \wedge Q_J  \wedge K^{J}{}_{I}  \nonumber\\
			&&- K^I{}_4 \wedge DQ_J  \wedge K^{J}{}_{I} ,  \label{NewInv5}
		\end{eqnarray}
		are topological invariants. Notice that $I_4$ is a linear combination of the 4-form
		\begin{equation}
			DK^{I}{}_{J}\wedge DK^{J}{}_{I} + 2 K^{I}{}_{J}\wedge K^{J}{}_{K}\wedge R^{K}{}_{I},
		\end{equation}
		which was reported in~\cite{Montesinos2021} and the 4-form~\eqref{NewInv1} with $K^I{}_4=Q^I$. On the other hand, the 4-form $I_5$  is readily recognized as a generalization of the 4-form~\eqref{NewInv3}.  Indeed, by taking $K^I{}_4=Q^I$, it is direct to see that~\eqref{NewInv5} reduces to~\eqref{NewInv3}, $I_5=I_3$. Two particular cases of~\eqref{NewInv5} are
		\begin{enumerate}[(i)]
			\item Setting  $K^{I}{}_{4}\equiv e^I$, $Q^I\equiv \star(R^I{}_J \wedge e^J)$ and $K^I{}_J\equiv\omega^I{}_J - \Gamma^I{}_J$, we find that $I_5=(1/2)I_2$ with $I_2$ given by~\eqref{ExI2}. This can be verified by using $De^I=K^I{}_J \wedge e^J$ and $D^2e^I=R^I{}_J\wedge e^J$.
			\item Setting  $K^{I}{}_{4}\equiv \star[(\star R^I{}_J) \wedge e^J]$, $Q^I\equiv \star(R^I{}_J \wedge e^J)$ and $K^I{}_J\equiv T_{K}{}^I{}_J e^K$, the 4-form $I_5$ reads
			\begin{eqnarray}
					I_5&=& -\sigma \big\{ \mathfrak{R}^I{}_{K} e^K \wedge \star(R_{JL} \wedge e^L)  \wedge D(T_{M}{}^J{}_I e^M) \nonumber\\
					&&+ D\big(\mathfrak{R}^I{}_{K} e^K\big) \wedge \star(R_{JL} \wedge e^L)  \wedge T_{M}{}^J{}_I e^M  \nonumber\\
				&&- \mathfrak{R}^I{}_{K} e^K \wedge D[\star(R_{JL} \wedge e^L)]  \wedge T_{M}{}^J{}_I e^M \big\}.
			\end{eqnarray}
		It is direct to see that this 4-form vanish if there is no torsion, $De^I=0$.
		\end{enumerate}
			
		\subsection{Identity coming from~\eqref{Iden-IV}}
		
		Using~\eqref{conn1} and~\eqref{curv1}, the identity~\eqref{Iden-IV} can be written as
		\begin{align}
			&K^{I}{}_{J}\wedge K^{J}{}_{K}\wedge DK^{K}{}_{I} + DK_{IJ} \wedge K^I{}_4 \wedge K^{J}{}_4  \nonumber\\
			& \hspace{3mm} + 2 K_{IJ} \wedge K^I{}_4 \wedge DK^J{}_4 \nonumber\\
			&= d\left( \frac{1}{3} K^{I}{}_{J}\wedge K^{J}{}_{K}\wedge K^{K}{}_{I} + K_{IJ} \wedge K^I{}_4 \wedge K^J{}_4 \right). \label{Iden2-IV}
		\end{align}
		Notice that~\eqref{Iden2-IV} does not depend on the parameter $l$, and hence no additional identities emerge. Then, directly from~\eqref{Iden2-IV} it follows that the integral of the 4-form
		\begin{eqnarray}
			I_6  &\equiv&  K^{I}{}_{J}\wedge K^{J}{}_{K}\wedge DK^{K}{}_{I}  + DK_{IJ} \wedge K^I{}_4 \wedge K^{J}{}_4  \nonumber\\
			&& + 2 K_{IJ} \wedge K^I{}_4 \wedge DK^J{}_4
		\end{eqnarray}
		is a topological invariant. The 4-form $I_6$ turns out to be a linear combination of the 4-form
		\begin{equation}
			K^{I}{}_{J}\wedge K^{J}{}_{K}\wedge DK^{K}{}_{I},
		\end{equation}
		 which was reported in~\cite{Montesinos2021} and the 4-form~\eqref{NewInv3} with $Q^I=K^I{}_4$; therefore, no new 4-form whose integral leads to new topological invariant arises in this case. 
		
		\vspace{3mm}
		
		\section{Conclusions}
		
		In this paper we have reported generalizations of the Nieh-Yan topological invariant in four dimensions. We have followed a systematic method along the lines of Refs.~\cite{Nieh-Yan1982,Zanelli1997,Nieh2018,Montesinos2021}, obtaining a family of 4-forms resulting from cases A, B, and C of Sec. \ref{sec-general}. In each case, explicit expressions for some of the resulting invariants are given and we showed how the Nieh-Yan form is obtained from particular choices of the 1-forms $Q^I$, $K^{I}{}_{4}$, and $K^I{}_J$. We have considered all the possible cases that emerge from the method, namely cases A, B, C, and D of Sec. \ref{sec-general}; this despite the fact that case~D turned out to be very simple and does not give rise to new topological invariants.
		
	    Regarding the invariant~\eqref{exampleA1}, it is worth noting that $f$ can be a functional depending on $e^I$ and $\omega^I{}_J$ through $e^I$  itself, $R^I{}_J$, and $De^I$. In this way, for instance, $f$ could be a function of the scalar curvature $ \mathfrak{R}= \mathfrak{R}^I{}_I $, i.e., $f=f(\mathfrak{R})$, or it could be a function of $\tau_I \tau^I$ where $\tau_I=T^J{}_{IJ}$, i.e., $f=f(\tau_I \tau^I)$. Additionally, in those cases where a scalar field $\phi$ is coupled to gravity, $f$ could be a function of the scalar field, i.e., $f=f(\phi)$. Furthermore, it would be also interesting to compute the topological invariants presented here in theories in which torsion emerges, as is the case of general relativity coupled to spin matter fields~\cite{Freidel2005,Perez-Rovelli-2006,Mercuri2006,Alexandrov_2008,Romero_2021} and  $f(\mathfrak{R})$ gravity with torsion~\cite{Rubilar_1998,Capozziello2010,MontesinosfR_2020}.
		
		Lastly, another interesting aspect to investigate is the addition of these topological invariants to actions for general relativity such as Palatini or Holst~\cite{Holst9605} actions and also to actions of alternative theories of gravity, and to study the implications of this fact both at Lagrangian and Hamiltonian levels. In fact, at the Hamiltonian level, it would be expected that these topological invariants would induce canonical transformations on the phase space~\cite{Perez2009}, which would introduce new phase space variables that could have geometric or physical relevance.
		
		\begin{acknowledgments}
			We thank Mariano Celada and Jorge Romero for carefully reading the manuscript and for their fruitful comments. This work was partially supported by Fondo SEP-Cinvestav and by Consejo Nacional de Ciencia y Tecnolog\'{i}a (CONACyT), M\'{e}xico, Grant No.~A1-S-7701.
		\end{acknowledgments}
		
		\bibliography{references}
	
\end{document}